\documentclass{iaus}
\usepackage{graphicx}

\title[Estimating the binary fraction of planetary nebula central stars]{Estimating the binary fraction of planetary nebula central stars} 

\author[D. Douchin et al.]{D. Douchin$^{1,2,3}$, O. De Marco$^{1,2}$, D.J. Frew$^{1,2}$, G. H. Jacoby$^4$, J.-C. Passy$^{5,11}$, T. Hillwig$^6$, S. B. Howell$^7$, H. Bond$^8$, A. Peyaud$^1$, A. Zijlstra$^9$, R. Napiwotzki$^{10}$, G. Jasniewicz$^3$, Q. Parker$^{1,2}$}

\affiliation{$^1$Macquarie University Research Centre in Astronomy, Astrophysics \& Astrophotonics $^2$Department of Physics \& Astronomy, Macquarie University Australia $^3$GRAAL, Universit\'e Montpellier 2, France $^4$Giant Magellan Telescope,USA $^5$University of Victoria, Canada $^6$Valparaiso University,  USA  $^7$NASA Ames Research Center, USA  $^8$Space Telescope Science Institute, USA      $^9$University of Manchester, UK $^{10}$University of Hertfordshire, UK   $^{11}$American Museum of Natural History, New York City, USA}

\pubyear{2011}
\volume{283}  
\pagerange{119--126}
\jname{Planetary Nebulae: an Eye to the Future}
\editors{Arturo Manchado, Letizia Stanghellini \& Detlef Sch\"onberner, eds.}
\begin{document}

\maketitle


\begin{abstract}

During the past 20 years, the idea that non-spherical planetary nebulae (PN) may need a binary or planetary interaction to be shaped was discussed by various authors. It is now generally agreed that the varied morphologies of PN cannot be fully explained solely by single star evolution. Observationally, more binary central stars of planetary nebulae (CSPN) have been discovered, opening new possibilities to understand the connections between binarity and morphology. So far, $\simeq$45 binary CSPN have been detected, most being close systems detected via flux variability. To determine the PN binary fraction, one needs a method to detect wider binaries. We present here recent results obtained with the various techniques described, concentrating on binary infrared excess observations aimed at detecting binaries of any separation.
\end{abstract}



\section{Introduction}

There is no general concensus yet about what shapes planetary nebulae (PNe). The idea that the presence of a stellar or substellar companion is needed to break the spherical symmetry and account for the observed geometries, called the Binary Hypothesis (see \cite{DeMarcoPASP}), is still being tested. The fact that 80\%  of PNe are non-spherical (\cite{nonspherical}) potentially implies that the proportion of binary central stars of PNe (CSPN) is much larger than current estimates, therefore indicating that PNe could be preferentially created via a binary channel. If the observed  fraction of binary CSPN were higher than expected in the current paradigm (35\% for binaries with separations $<500$ AU), this would support the idea that PNe are preferentially a binary phenomenon. To estimate the binary fraction, we need an unbiased sample of PNe. The determination of whether a CSPN is binary is not trivial and requires a suite of methods, which we describe below. We also present new obervational results and how this constrains the binary fraction. We conclude with some thoughts to refine this value.

 
 

\section{Methods to investigate binarity}

We here present the methods used to investigate binarity in our PN sample: flux and spectral variability and infrared (IR) excess, as well as recent results obtained with them.

The binary-induced flux variability method is based on the variation of brightness of the binary system. The three main causes of this flux variability are eclipses, tidal deformations, but mainly irradiation effects from the hot companion onto the cold one create flux changes. 
This method gives a close binary fraction of about 15-20\% (see \cite{Bond2000}, \cite{Miszalski2009}). It is biased to short periods as the three causes of binary induced flux variation all diminish with increasing separation. It is also biased to intrinsically faint stars to avoid periodic variability caused by stellar winds. Pulsations can also be accountable for flux variation. New discoveries will be possible thanks to Kepler, whose field includes 6 PNe, including Kronberger 61, newly imaged by us in March 2011 (see http://www.gemini.edu/node/11656). Five of the six objects show flux variability (see \cite{douchin2010}.)

Spectroscopic variability is used to detect the motion of binary systems around their barycenter. Binaries with periods up to a month could potentially be detected with this method. Intrinsically faint stars, with weak or no wind, are targets of choice. This introduces a bias and imposes serious technical restrictions to obtain a sufficient signal-to-noise ratio, ranking this method down for statistical purposes required in our case. Out of 7 central stars targeted by the VLT/UVES, we found that one, A 14, has a radial velocity (RV) variability confirmed at a 3$\sigma$ level. Another 3 central stars are likely to be binaries (Douchin et al., in preparation).

The IR excess technique detects the IR emission from a cool companion using high precision optical and NIR photometry. So far, the detection limit is an M8-type star for intrinsically faint central stars (Mv = 6-8). This technique has the advantage that it is less biased by binary separation. We aim to perform high precision photometry for the entire 2 kpc volume-limited sample of Frew (\cite{Frew2008}; \cite{FrewParker2007}). So far, out of 28 central stars of PNe that were observed in the I-band, we have solidly detected an I-band excess in 3, with another possible 5 detections. The spectral types for the 3 detected companions are likely in the early M regime. We have not yet quantified our survey's bias against fainter companions (Passay et al., in preparation). 

\vspace{-.4 cm}

\section{Current and future projects}


 
We are currently analysing additional I and J-band photometry datasets which will provide us a preliminary estimate of the binary fraction (separations less than 500 AU).

In addition, space-based  observations such as GAIA have a potential capability to detect the reflex orbital motion of central stars of PNe for binaries with intermediate separations, and the extreme photometric precision of Kepler can be used to gauge how many close binaries have avoided detection in previous ground based surveys (see \cite{douchin2010}). Also, future ELTs will have the resolving power and sensitivity to detect companions via AO-enhanced direct imaging at separations of a few AU and larger out to 2 kpc, to provide good complementarity to current methods.

\end{document}